\documentclass[preprint,showpacs,preprintnumbers,amsmath,amssymb]{revtex4}

\usepackage{graphicx}
\usepackage{dcolumn}
\usepackage{bm}
\usepackage{epsfig}
\newcommand{\be}{\begin{eqnarray}}
\newcommand{\ee}{\end{eqnarray}}

\begin{document}


\title{Magnetic ordering and quantum statistical effects in strongly
repulsive  Fermi-Fermi and Bose-Fermi mixtures}

\author{Xi-Wen Guan$^{\dagger}$, Murray T Batchelor$^{\dagger \ddagger}$
  and Jen-Yee Lee$^{\dagger}$}
\affiliation{${}^{\dagger}$ Department of Theoretical Physics, Research School of Physical Sciences and Engineering,\\
Australian National University, Canberra ACT 0200,  Australia\\
${}^{\ddagger}$ Mathematical Sciences Institute,\\
Australian National University, Canberra ACT 0200,  Australia}

\date{\today}

\begin{abstract}
\noindent
We investigate magnetic properties and statistical effects in 1D
strongly repulsive two-component fermions and in a 1D mixture of 
strongly repulsive polarized fermions and bosons.  
Universality in the characteristics of phase transitions, magnetization and 
susceptibility in the presence of an external magnetic field $H$ are analyzed 
from the exact thermodynamic Bethe ansatz solution. 
We show explicitly that polarized fermions with a repulsive interaction have 
antiferromagnetic behavior at zero temperature. 
A universality class of linear field-dependent magnetization persists for
weak and finite strong interaction. 
The system is fully polarized when the external field exceeds
the critical value $H^F_c\approx \frac{8}{\gamma}E_F$, where $E_F$ is the Fermi energy 
and $\gamma$ is the dimensionless interaction strength. 
In contrast, the mixture of polarized fermions and bosons in an
external field exhibits square-root field-dependent magnetization in
the vicinities of $H=0$ and the critical value $H=H^M_c\approx
\frac{16}{\gamma}E_F$.
We find that a pure boson phase occurs in the absence of the external field, 
fully-polarized fermions and bosons coexist for $0<H<H^M_c$, and a fully-polarized fermion phase
occurs for  $H\ge H_c^M$. 
This phase diagram for the Bose-Fermi mixture is reminiscent of
weakly attractive fermions with population imbalance, where the
interacting fermions with opposite spins form singlet pairs.

\end{abstract}

\pacs{03.75.Ss, 05.30.Fk, 71.10.Pm}

\keywords{}

\maketitle 
\section{Introduction}

The recent success in experimentally realizing degenerate quantum gases in low
dimensions \cite{E-TG1,E-TG2,E-TG3,E-TG4,E-TG5} has revived 
interest in one-dimensional (1D) integrable models of interacting
fermions and bosons \cite{1D-1,1D-2,1D-3,1D-4,Li,Zhou,Gia,Zoellner}.
The 1D atomic gases with internal degrees of freedom are tunable
interacting many-body systems featuring novel strong correlation
effects and subtle quantum phase transitions.  
Exotic quantum phases in 1D two-component attractive fermions have 
received considerable interest
\cite{Fuchs,Tokatly,Astr,Orso,Hu,GBLB,Feiguin,Wadati,Ueda,Batrouni,Parish,Ying,cazalilla2,Liu,Casula}
due to the experimental observation of fermionic superfluidity and phase
transitions \cite{I-F1,I-F2,I-F3}. 
For certain regimes, the two-component
Fermi gas with population imbalance can be viewed as a strongly
interacting Bose-Fermi mixture \cite{Ketterle,Recati}. 
However, subtle differences between the bosonic dimer and pure bosons have been
observed experimentally \cite{Md1,Md2}.  
For repulsive interaction,
the two-component Fermi gas exhibits antiferromagnetic behavior
\cite{Fuchs,Tokatly,MC,BBGO,Wadati2,Snoek}.  
In contrast to the two-component Fermi gas, the 1D spinor Bose gases
\cite{Li,Fuchs2,BBGO2,GBT} possesses novel ferromagnetic properties and
collective dynamics of spin waves at temperatures below the degenerate temperature. 
The subtlety in 1D quantum many-body physics \cite{Giamarchi} is a result of 
the dynamical interaction between identical particles and their statistics.

On the other hand, the recent success in tuning interspecies Feshbach resonances in
Bose-Fermi mixtures of ultracold atoms \cite{Md1,Md2,Dick}, such as
$^{6}$Li+$^{7}$Li, $^{6}$Li+$^{23}$Na, $^{40}$K+$^{87}$Rb,
$^{6}$Li+$^{87}$Rb, opens up a new gateway for exploring many-body physics,
including superfluids and Mott insulators, spin and charge density
waves, phase separation, the BCS-BEC crossover etc.  
On the theoretical side, Bose-Fermi mixtures have been studied through
various techniques like the mean-field approximation \cite{MF},
Luttinger liquid formalism \cite{LL,Mathey}, Quantum Monte Carlo \cite{Zujev}, 
bosonization techniques \cite{BT1,BT2}, exact
solutions using the Bethe ansatz \cite{BA1,BA2} and other methods
\cite{Other}. 
However, the role of quantum fluctuations is enhanced in 1D 
compared to the three-dimensional case to the extent that 
traditional mean-field theories fail for strong interaction in 1D. 
The exact Bethe ansatz solution of 1D many-body systems provides more reliable
physics than the mean-field theory.

In this paper, we investigate external field-dependent magnetic properties and 
statistical effects of 1D two-component
fermions with repulsion and a 1D mixture of polarized fermions and
bosons by means of their exact thermodynamic Bethe ansatz solution.  
The antiferromagnetic groundstate properties and critical
behavior of two-component repulsive fermions are studied in detail. 
The universality class of linear field-dependent behavior of
magnetization is predicted for weak and finitely strong coupling regimes. 
However,  in the Tonks-Girardeau limit, a van Hove
singularity in the susceptibility occurs as the external field
approaches a critical field. In this limit, the system becomes a
paramagnet, i.e., a system with  free spins.  
However, the existence of bosons in the Bose-Fermi-Fermi mixture destroys the antiferromagnetic
ordering in the two-component interacting fermions.  In the absence of
the external field, the state of pure bosons is among the groundstates. 
When the external field is turned on, a second order phase
transition from a pure boson phase into a mixed boson-fermion phase occurs. 
A fully-polarized Fermi liquid occurs when the external field exceeds a critical value.
We calculate the explicit details of these transitions.

This paper is set out as follows. In section \ref{Hamiltonian}, we
recall the Bethe ansatz solution of the 1D integrable model of mixed bosons and fermions. 
In section \ref{Fermi}, we study the antiferromagnetic behavior of the
1D strongly repulsive interacting Fermi gas with polarization. The
phase diagram of the model is presented.  The quantum phase
transitions and magnetic properties of the Bose-Fermi mixture are
studied in section \ref{Mixture}.
Section \ref{Conclusion} is devoted to concluding remarks and a brief
discussion.

\section{ The Hamiltonian}
\label{Hamiltonian}

We consider a $\delta$-function interacting system of $N$ bosons and
fermions with equal mass $m$ and internal degrees of freedom, or a
mixture of two-component fermions and spinless bosons, constrained by
periodic boundary conditions to a line of length $L$ in an 
external magnetic field $H$.
The Hamiltonian \cite{Sutherland,Lai-Yang} is
\begin{eqnarray}
{H}&=&-\frac{\hbar ^2}{2m}\sum_{i = 1}^{N}\frac{\partial
^2}{\partial x_i^2}+\,g_{\rm 1D} \sum_{1\leq i<j\leq N} \delta
(x_i-x_j)\nonumber\\
&&-\frac12 H(M_1-M_2).
\label{Ham-1}
\end{eqnarray}
The quantum numbers $M_1$ and $M_2$ are the
numbers of fermions with spin-up and spin-down, respectively. 
Under exchange  of  spatial and internal spin coordinates between two particles 
the wavefunctions of the Hamiltonian (\ref{Ham-1}) are symmetric for 
bosons or for fermions with opposite hyperfine states and
antisymmetric for fermions with the same hyperfine states.

For maintaining the integrability of the model, we consider the
same mass for bosons and fermions and the same interaction strength $g_{\rm 1D}$
between bosons, between bosons and fermions and between fermions with
opposite hyperfine states. 
The interaction is attractive for $g_{\rm 1D}<0$ and repulsive for $g_{\rm 1D}>0$.  
There is no $\delta$-type interaction between fermions with the same hyperfine states. 
Although these conditions appear somewhat restrictive, the model still captures 
the essential physics \cite{BA1} relevant to the current experimental
observations \cite{Ketterle,Md1,Md2}. 
The interspecies interaction can be tuned from strongly attractive ($g_{\rm 1D}\rightarrow -\infty$) to
strongly repulsive ($g_{\rm 1D} \rightarrow +\infty$) via Feshbach
resonances and optical confinement.  The coupling constant $g_{\rm
1D}$ can be written in terms of the scattering strength $c={2}/{a_{\rm
1D}}$ as $g_{\rm 1D} ={\hbar ^2 c}/{m}$.
In principle an effective $1$D scattering length $a_{\rm 1D}$ can be expressed through the 
$3$D scattering length for the bosons and fermions confined in a $1$D geometry. %
We use the dimensionless coupling constant $\gamma={mg_{\rm 1D}}/{(\hbar^2n})$ 
for physical analysis. 
Here $n={N}/{L}$ is the linear density.

This model was solved in 1963 by Lieb and Liniger for the special case of spinless bosons \cite{L-L}. 
Experimental realization of 1D interacting Bose gases \cite{E-TG1,E-TG2,E-TG3,E-TG4} 
has stimulated further interest in various integrable models, including 
interacting two-component fermions solved by Yang and Gaudin \cite{Yang,Gaudin}, 
the Bose-Fermi mixture solved by Lai and Yang \cite{Lai-Yang}, and multi-component bosons and fermions
solved by Sutherland \cite{Sutherland}.

The Bethe ansatz equations for the Bose-Fermi mixture (\ref{Ham-1}) with
an irreducible representation $[2+M_{b}, 2^{M_{2}-1}, 1^{M_{1}-M_{2}}] $ are \cite{Lai-Yang}
\begin{eqnarray}
\exp(\textrm{i}k_{j}L)&=&\prod_{\alpha=1}^{M} \frac{k_{j}-\lambda_{\alpha}+\textrm{i}c'}{k_{j}-\lambda_{\alpha}-\textrm{i}c'}\nonumber\\
\prod_{j=1}^{N} \frac{\lambda_{\alpha}-k_{j}+\textrm{i}c'}{\lambda_{\alpha}-k_{j}-\textrm{i}c'} 
&=& -\prod_{\beta=1}^{M}
\frac{\lambda_{\alpha}-\lambda_{\beta}+\textrm{i}c}{\lambda_{\alpha}-\lambda_{\beta}-\textrm{i}c}\nonumber\\
&&\times \prod_{b=1}^{M_{b}}
\frac{\lambda_{\alpha}-A_{b}-\textrm{i}c'}{\lambda_{\alpha}-A_{b}+\textrm{i}c'}\label{BA-M}\\ 
\prod_{k=1}^{M} \frac{A_{b}-\lambda_{k}-\textrm{i}c'}{A_{b}-\lambda_{k}+\textrm{i}c'} &=&1.\nonumber
\end{eqnarray}
We define $M=M_2+M_b$, where $M_b$ is the number of bosons.  
In these equations $k_j$, with $j=1,\ldots, N$, are the quasimomenta of the
particles and $\lambda_{\alpha}$, with $\alpha=1,\ldots, M$, are parameters for the
fermions with spin-down and bosons. $A_b$ with $b=1,\ldots, M_b$ are
the parameters for the bosons.  
In the Bethe ansatz process \cite{Lai-Yang}, the conjugate representation 
$[M_{1}, M_{2},1^{2+M_{b}}]$ was used to solve matrix eigenvalue
equations associated with the irreducible representation
$[2+M_{b}, 2^{M_{2}-1}, 1^{M_{1}-M_{2}}] $ following Sutherland's paper \cite{Sutherland}.

The Bethe ansatz solution of the Bose-Fermi mixture was considered with 
renewed interest \cite{BT2,BA1,BA2} due to the experimental observation of new
quantum phases \cite{Md1,Md2}. 
The groundstate properties, correlation functions and harmonic trapping for the mixture of
fully-polarized fermions and bosons were studied in Ref. \cite{BA1}. 
The model of arbitrarily polarized fermions and spinless bosons was  investigated in
the weak coupling regime \cite{BA2}. The Bethe ansatz equations
(\ref{BA-M}) also contain those for two-component fermions \cite{Yang} when $M_b=0$. 
However, the magnetic properties and
phase transitions for two-component fermions in the repulsive regime 
and the mixture of polarized fermions and bosons have not been
comprehensively investigated. 
We turn now to the investigation of their magnetic properties and phase transitons in the
repulsive regime.

\section{Two-component fermions}
\label{Fermi}

It is well known that the groundstate for 1D interacting fermions is antiferromagnetic. 
This is proved to be the case by the Lieb-Mattis theorem \cite{LiebMattis}. 
They showed that for a system with total spin $S'>S$, the lowest energy belonging to that
system is also greater, $E(S')>E(S)$. 
Hence when there is no external field, the system is not polarized. 
Another way of showing this is from the Pauli exclusion principle. 
When there is no interaction at all, the groundstate can only have one pair of fermions 
(spin-up and spin-down) having a particular pseudomomentum $k$. When a
repulsive interaction $\gamma$  is ``switched'' on, the pair will split so
that no two fermions can have the same value for $k$. 
However, the sequence of spin-up and spin-down fermions is still unaltered, which
shows that they remain anti-ferromagnetic even when $\gamma>0$.
The magnetization of the two-component fermion system in  the weak coupling limit,
$\gamma \ll 1$, can be directly obtained from the groundstate energy
derived from the discrete BA equations \cite{BBGO}
\begin{equation}
\frac{E}{L}\approx \frac{c}{2}n^2(1-P^2)+\frac{\pi^2}{12}n^3+\frac{\pi^2}{4}n^3P^2
\end{equation}
where in an obvious notation the polarization $P=(N_{\uparrow}-N_{\downarrow})/{N}$.
Defining the magnetization per particle as $m^z={nP}/{2}$, the
magnetization has a linear field-dependent form given by
\begin{equation}
H\approx 2n\left(\pi^2 -2\gamma \right) m^z.\label{F-w-mz}
\end{equation}
When the external field exceeds the critical value $H_c^F\approx n^2(\pi^2-2\gamma)$ 
a fully-polarized phase occurs.
The magnetic susceptibility  is $\chi \approx 1/2n(\pi^2-2\gamma)$ for finite $H$. 
However, from field theory, $\chi \approx 1/(2\pi v_s)$ at $H=0$. 
Here the spin velocity is $v_s\approx n\pi(1-\gamma/\pi^2)$ \cite{Giamarchi,Fuchs,Michael}.

We turn now to the magnetic properties  of the $1$D
exactly solved model of  two-component fermions with arbitrary polarization in the 
strong coupling regime $\gamma \gg 1$.  
In the thermodynamic limit, $L,N \to \infty$ with $N/L$ finite, the Bethe ansatz equations 
can be written in terms of the spin and charge densities in the form \cite{Yang,Gaudin}
\begin{eqnarray}
\rho(k)+\rho^h(k)&=&\frac{1}{2\pi}+\frac{1}{2\pi} \int_{-B}^{B}\frac{c\sigma(\lambda)}{c^2/4+(k-\lambda)^2}d\lambda\\
\sigma(\lambda)+\sigma^h(\lambda)&=&\frac{1}{2\pi}\int_{-Q}^{Q}\frac{c
  \rho(k)}{c^2/4+(\lambda-k)^2}dk \nonumber\\
&&-\frac{1}{2\pi} \int_{-B}^{B}\frac{2c\sigma(\lambda)}{c^2+(\lambda-\lambda')^2}d\lambda'.\label{BA-F}
\end{eqnarray}
The integration limits $Q$ and $B$ are determined by
the total number of particles 
$N/L=\int_{-Q}^Q\rho(k)dk$ and the number of  spin-down fermions 
$M_2/L=\int_{-B}^{B}\sigma(\lambda)d\lambda$.  
For the groundstate $\rho^h(k)=\sigma^h(\lambda)=0$, i.e. there are no holes for both
charge and spin rapidities.

The groundstate properties can be determined from 
the TBA equations in the limit $T\to 0$ \cite{Lai-1,Takahashi}.
In terms of the dressed energies $\epsilon(k):=T\ln[\rho^h(k)/\rho(k)]$ for the charge and 
$\phi(\lambda):=T\ln[\sigma^h(\lambda)/\sigma(\lambda)]$ for the spin degrees of freedom, 
these equations are
\begin{eqnarray}
\epsilon(k)&=&k^2-\mu  -\frac12{H}+\frac{1}{2\pi}\int_{-B}^{B}\frac{c\phi^-(\lambda)}{c^2/4+(k-\lambda)^2}d\lambda
\label{TBA-F1}\\
\phi(\lambda)&=&g_1(\lambda) -\frac{1}{2\pi}\int_{-
  B}^{B}\frac{2c\phi^-(\lambda')}{c^2+(\lambda-\lambda')^2}d\lambda'
\label{TBA-F} 
\end{eqnarray} 
where the driving term is given by
\begin{equation}
g_1(\lambda)=H+\frac{1}{2\pi}\int_{-Q}^Q\frac{c\epsilon^-(k)}{c^2/4+(\lambda-k)^2}dk. 
\end{equation}
The negative part of the dressed energy, $\epsilon(k)\le 0$ for $|k|\le Q$, or $\phi(\lambda)\le 0$
for $|\lambda|\le B$, corresponds to occupied states, with the 
positive part of $\epsilon(k)$ and $\phi(\lambda)$ corresponding to unoccupied states.  
We clearly see from the TBA equations (\ref{TBA-F1}) and (\ref{TBA-F}) that the
spin interaction is antiferromagnetic with an effective spin-spin
exchange interaction depending on the energy of the system.  For
$H=0$, the driving term gives rise to an asymptotic condition $\phi
(\infty ) =0$ below the $\lambda$-axis. This leads to a maximum
$\phi^-(\lambda)$ which gives the lowest energy state of the system.
For strong coupling $\gamma \gg 1$, 
the driving terms in the second TBA equation  (\ref{TBA-F}) become
$g_1(\lambda) \approx H-cP_0/(c^2/4+\lambda^2)$ where we ignore
$O(1/c^2)$ contributions.  Here
$P_0=-\frac{1}{2\pi}\int_{-Q}^{Q}\epsilon^-(k)dk$ is the pressure.
The driving term $g_1$ clearly indicates that the form of 
equation (\ref{TBA-F}) indicates an effective antiferromagnetic spin-spin
exchange interaction with an effective coupling constant $J=-2P_0/c<0$
in contrast to the effective  ferromagnetic coupling in spinor Bose gas \cite{GBT}, for which $J=2P_0/c$. 
For the groundstate we find that
$\phi_0(\lambda)=-\frac{\pi P_0}{c\,\cosh(\pi \lambda/c)}$. 
On solving the TBA equation (\ref{TBA-F}), the leading behavior for the energy per unit length and the chemical
potential is given by 
\begin{eqnarray}
F(0)/L&=&\frac{1}{3}n^3\pi^2 \left(1-\frac{4\ln 2}{\gamma}\right) \\
 \mu&=&n^2\pi^2 \left(1-\frac{16\ln 2}{3\gamma}\right)
\end{eqnarray} 
which coincide with previous results \cite{Fuchs}. 

\subsection{Finite external field $H$}

When the external field is applied, the spin-up states are energetically favoured due to Zeeman splitting.  
As the magnetic field increases, the Fermi sea of the spin sector $\phi(\lambda)$ is raised. 
For small field, $H \ll 1$, the second TBA equation (\ref{TBA-F}) becomes 
\begin{eqnarray}
\phi(\lambda)&=&\frac12 {H}-\frac{2\pi
  P}{2c\cosh(\pi\lambda/c)}\nonumber\\
& &+\int_{-\infty}^{\infty}G(\lambda-\lambda') \phi^{+}(\lambda')d\lambda'\label{WH-1}
\end{eqnarray}
with the function
\begin{equation}
G(\lambda)=\frac{1}{2\pi}\int_{-\infty}^{\infty}\frac{1}{1+{\rm e}^{|\omega c|}}{\rm e}^{-\textrm{i}\omega \lambda}d\omega.
\end{equation}
Making a change of variables $\lambda\rightarrow\lambda+B$ and
defining $y(\lambda)\equiv\phi(\lambda+B)$, equation (\ref{WH-1}) becomes
\begin{eqnarray}
y(\lambda) &=& \frac12 {H}-\frac{2\pi
P}{2c\cosh(\pi(\lambda+B)/c)}+\int_{0}^{\infty}G(\lambda-k)y(k)dk\nonumber\\
& &+\int_{0}^{\infty}G(\lambda+2B+k)y(k)dk.
\end{eqnarray}
Observing that $G(\lambda+2B+k)\rightarrow {\rm e}^{-2\pi B/c}$ for $B\gg 1$, 
we can take the expansion $y=y_{1}+y_{2}+\ldots$ with respect to
the order of ${\rm e}^{-2\pi mB/c}$ where $m=1,2,\ldots$. It follows that
\begin{eqnarray}
&& y_{1}(\lambda)+y_{2}(\lambda)+\ldots =
\frac12{H}-{2\pi P}{c^{-1}}{\rm e}^{-\pi\lambda/c}{\rm e}^{-\pi B/c} \nonumber \\ 
&& +\int_{0}^{\infty}G(\lambda+2B+k)(y_{1}(k)+y_{2}(k)+\ldots)dk \nonumber \\
&& +\int_{0}^{\infty}G(\lambda-k)(y_{1}(k)+y_{2}(k)+\ldots)dk.
\end{eqnarray}
Collecting terms of the same order then gives
\begin{eqnarray}
y_{1}(\lambda)&=&\frac12{H}-{2\pi P}{c^{-1}}{\rm e}^{-\pi\lambda/c}{\rm e}^{-\pi B/c}
+\int_{0}^{\infty}G(\lambda-k)y_{1}(k)dk
\\
y_{2}(\lambda)&=&\int_{0}^{\infty}G(\lambda+2B+k)y_{1}(k)dk+\int_{0}^{\infty}G(\lambda-k)y_{2}(k)dk
\end{eqnarray}
and so on. To a sufficient approximation, we only need to consider the equation for $y_{1}(\lambda)$. 
In considering the leading contribution in powers of the exponential
$G(\lambda+2B+k)\rightarrow {\rm e}^{-2\pi B/c}$ for $B\gg 1$, equation
(\ref{WH-1})  reduces to the standard Wiener-Hopf equation (for the 
Wiener-Hopf technique, see, e.g., \cite{Takahashi,Hubbardbook,Schlot,BGOT,Lee-t}). 
After tedious calculation we obtain the free energy, pressure and chemical
potential as
\begin{eqnarray}
& &F\approx\frac{1}{3}\pi^{2}n^{3}\left(1-\frac{4\ln
2}{\gamma}\right)-\frac{3H^{2}\gamma}{8\pi^{4}n}\left(1+\frac{12 \ln
2}{\gamma}\right)\\
& &P\approx\frac{2}{3}\pi^{2}n^{3}\left(1-\frac{6\ln
2}{\gamma}\right)+\frac{9H^{2}\gamma}{8\pi^{4}n}\left(1+\frac{104\ln
2}{9\gamma}\right)\\
& & \mu\approx\pi^{2}n^{2}\left(1-\frac{16\ln
2}{3\gamma}\right)+\frac{3H^{2}\gamma}{4\pi^{4}n^{2}}\left(1+\frac{34\ln
2}{3\gamma}\right).
\end{eqnarray}
Moreover, the magnetization $m^{z}(H)$ and
susceptibility $\chi(H)$ follow in the small field regime using the 
formula $m^{z}=-\partial F(H)/\partial H$ and $\chi(H)=\partial m^{z}/\partial H$. 
Hence
\begin{eqnarray}
m^{z}(H)&\approx&\frac{3H\gamma}{4\pi^{4}n^2}\left(1+\frac{12\ln
2}{\gamma}\right)\label{F-mz}
\\
\chi(H)&\approx&\frac{3\gamma}{4\pi^{^4}n^2}\left(1+\frac{12\ln
2}{\gamma}\right).\label{F-chi}
\end{eqnarray}

\begin{figure}
\includegraphics[width=0.82\textwidth]{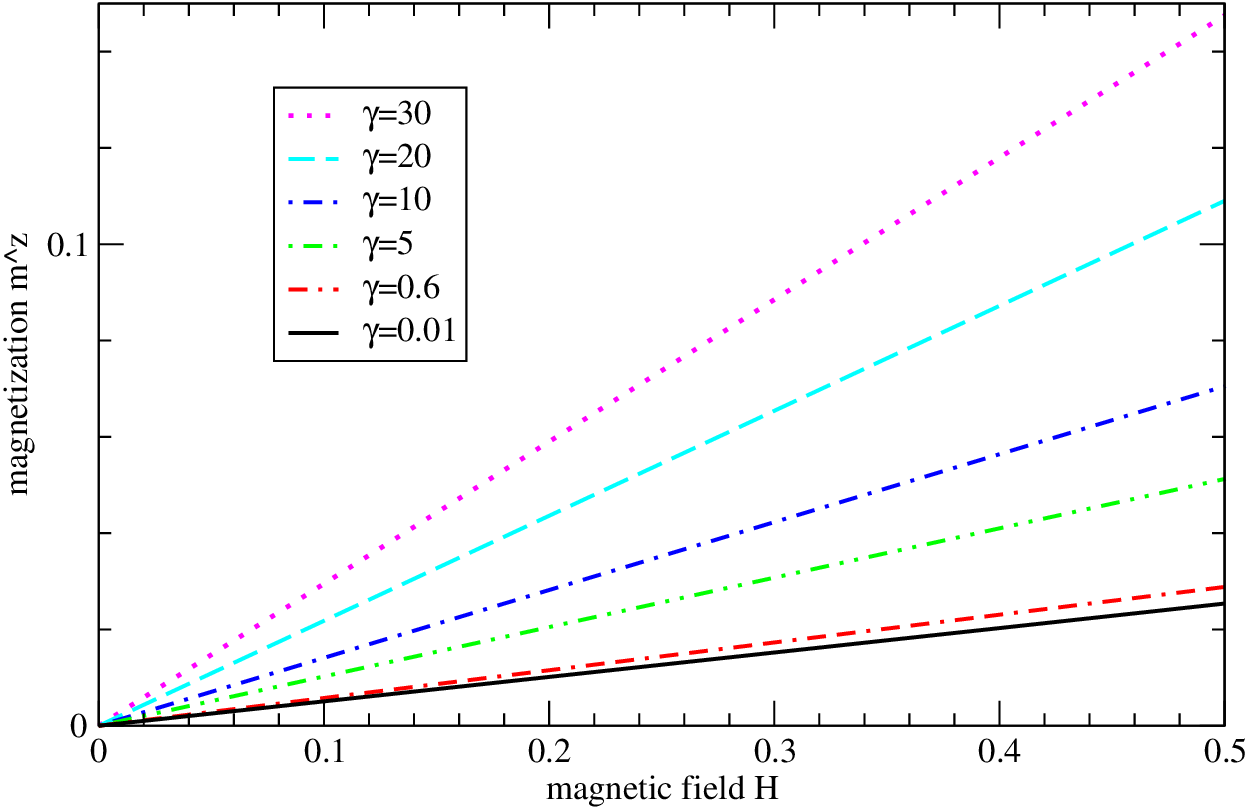}
\caption{Magnetization $m^{z}(H)$ vs external field $H$ for $n=1$ and
  different values of interaction strength $\gamma$. The
  small field magnetization curves are plotted from equations (\ref{F-w-mz}) and
  (\ref{F-mz}) for the weak and finitely strong coupling regimes, respectively.}  
\label{fig:F-mz}
\end{figure}

In the small field regime, the higher-order logarithmic correction to the
susceptibility can be calculated in the same fashion by solving to 
next order in the Wiener-Hopf equations.  
Thus finite logarithmic singularities are expected in the susceptibility \cite{Takahashi,Schlot}.  
As $n\to 0$ the susceptibility diverges as a
consequence of the van Hove singularity of the empty band. 
The linear field-dependent behavior of the magnetization is maintained for
finitely strong and weak repulsion, see figure \ref{fig:F-mz}.  
However, for the Tonks-Girardeau limit, $\gamma\to \infty$, 
caution should be paid to the critical behavior of the polarized fermions. 
The spin velocity is proportional to the inverse of the susceptibility. 
Therefore, in this limit the spin velocity $v_s$
tends to zero such that the spin propagation is almost frozen. 
As a consequence the statistical effect is fully suppressed and the
Tonks-Girardeau Fermi gas becomes a paramagnet or say free spins. 
Therefore, we expect an infinitely divergent susceptibility to occur for the system in the limit $\gamma \to \infty $. 
In the next part, we shall search for evidence of  the paramagnetic behavior
in the Tonks-Girardeau Fermi gas with population imbalance.

If the external magnetic field is sufficiently strong, the system can be fully-polarized.  
{}From the the TBA equations (\ref{TBA-F1}) and (\ref{TBA-F}), we find that if the
field $H \ge H_c^F=4 P_0/c $, then $\phi (0) \ge 0$. 
The critical value $H_c^F$ indicates a phase transition from a partially-polarized
ferromagnetic phase to a fully-polarized ferromagnetic phase. 
For the limit $H\to H_{c}^F$, the system is almost fully polarized, corresponding 
to a very small Fermi sea for the spin sector of the TBA equation.
The critical behavior may be directly derived from the TBA equations (\ref{TBA-F1}) and (\ref{TBA-F})
or the Bethe ansatz equations (\ref{BA-F}).  In the vicinity of
$H_{c}^F$, we can approximate the integral for the spin part by taking
the area under the curve as a rectangle as usually done in spin chains
and ladders \cite{BGOT}. 
In this way
\begin{eqnarray}
\phi(\lambda) &\approx&
H-\frac{Pc}{c^{2}/4+\lambda^{2}}-\frac{2Bc\phi^{-}(0)}{\pi(c^{2}+\lambda^{2})}\label{con-F}
\end{eqnarray}
with $\phi(0)\approx\frac{H-4P/c}{1+2B/\pi c}$. 
Further, $\phi(B)=0$ provides the value of the integration boundary, with
\begin{equation}
B=c\sqrt{\frac{4P/c-H}{5H-4P/c}}.
\end{equation}
On the other hand, from the Bethe ansatz equations (\ref{BA-F}), we
find
\begin{equation} 
B=\frac14{c\pi}(1-\sqrt{1-2\alpha })
\end{equation} 
if $\alpha =M/N \ll 1$. Here $M=M_2$ is the number of spin-down fermions.
Thus we have an explicit  relation 
\begin{equation}
\alpha =\frac{4}{\pi}\sqrt{(H_c^F-H)/(5H_c^F-H)}\label{alpha}
\end{equation}
between the numer of spin-down fermions and the external field, 
where $H_c^F=4P/c$.

Substituting the expression (\ref{con-F}) into the first equation of
the TBA (\ref{TBA-F}) and also using the conditions $\epsilon(Q)=0$ and
$\partial P/\partial \mu =n$, we obtain the leading terms for the pressure and the free energy
\begin{eqnarray}
P(H)&\approx&\frac{2}{3}\pi^{2}n^{3}\left(1-\frac{72}{\gamma}\frac{(H_{c}-H)^{1/2}}{\pi(5H-H_{c})^{1/2}}\right)
\\
F(H)&\approx&-\frac12{nH}+\frac{1}{3}\pi^{2}n^{3}-\frac{4n(H_{c}-H)^{3/2}}{\pi(5H-H_{c})^{1/2}}.
\end{eqnarray}
The groundstate energy 
\begin{equation}
\frac{E}{L}\approx \frac{1}{3}n^3\pi^2\left(1-\frac{8\alpha}{\gamma}\right)
\end{equation}
can be obtained from the relation $E/L=F(H)/L+Hn(1-2\alpha)/2$, 
which is consistent with the BA result \cite{BBGO}. 
Here recall that $\alpha =M/N$.
The critical field $H_{c}^F\approx 4P/c \approx\frac{8}{\gamma}E_F$, where 
$E_F=\frac{1}{3}n^2\pi^2$ is the Fermi energy.  
These explicit results indicate that for $\gamma \to \infty$, the system
becomes a paramagnet and statistical  interaction is suppressed. 
It also follows that the magnetization $m^{z}(H)$ and susceptibility $\chi(H)$
in the limit $ H\to H_{c}^F$ are given by
\begin{eqnarray}
m^{z}(H) &\approx&
\frac{n}{2}-\frac{6n(H_{c}-H)^{1/2}}{\pi(5H-H_{c})^{1/2}}-\frac{10n(H_{c}-H)^{3/2}}{\pi(5H-H_{c})^{3/2}}\\  
\chi(H)  &\approx &\frac{3n}{\pi(5H-H_{c})^{1/2}(H-H_{c})^{1/2}}\nonumber\\
&&+\frac{15n(H_{c}-H)^{1/2}}{\pi(5H-H_{c})^{3/2}}.
\end{eqnarray}
The expression for the susceptibility is seen to be divergent when $H \to H_{c}^F$. 
A quantum phase transition occurs as
the external field is greater than the critical field $H_c^F$, see the
phase diagram in  figure~\ref{fig:phase}.  The square-root behavior of
the magnetization is a  consequence of the van Hove singularity in one
dimension. 

The analytic results for the magnetic properties show that spin-spin
interaction for system of the polarized fermions with strong coupling $\gamma \gg 1$ 
can be effectively described by the isotropic spin-1/2 Heisenberg chain
with a weak antiferromagnetic coupling $J\approx -4E_F/\gamma$. 
For the isotropic spin-1/2 Heisenberg chain \cite{Takahashi}, the
susceptibility at $H=0$ is given by $\chi\approx 1/J\pi^2$ which
coincides with the result (\ref{F-chi}). 
Furthermore, we can find the spin velocity at $H=0$,
\begin{equation}
v_s \approx\frac{2\pi^3n}{3\gamma}\left(1-\frac{12\ln 2}{\gamma}\right).
\end{equation}
The antiferromagnetic spin-spin exchange interaction embeded in the 1D
interaction polarized fermions may provide insight into understanding
the spin segregation in a trapped Fermi gas \cite{Du}

\section{Bose-Fermi mixture}
\label{Mixture}

The earliest study of the groundstate of the 1D mixture of bosons and
polarized fermions dates back to Lai and Yang in 1971 \cite{Lai-Yang} who showed 
by numerically solving a set of coupled integral equations that 
the energy of the system is a monotonic decreasing function with respect to the ratio of
bosons and the total number of particles in the system. 
This implied that the groundstate is occupied only by bosons. 
The same signature was found recently by other  groups \cite{BT2,BA1,BA2}. 
The groundstate energy for the weak coupling regime was given explicitly in
Ref. \cite{BA2}. In addition, study of the quasimomentum distributions \cite{HuZhangLi2006} 
showed that the quasimomentum distribution is more compressed when there are no
fermions. 
As the number of fermions increases, the quasimomentum
distribution becomes more spread out due to the Fermi pressure.

It is thus natural to ask what the true physical groundstate is in the
presence of an external magnetic field.
What kind of magnetism  does the model exhibit? 
As far as we understand, the magnetic properties of the mixture of 
bosons and polarzied fermions have not been studied yet.  
It is of interest to know how the addition of
bosons influences the antiferromagnetic groundstate properties of the
spin-$\frac12$ fermionic system.  
This provides further insight in
understanding the signature of the strongly interacting
Bose-Fermi mixture and the two-component Fermi-Fermi mixture with
population imbalance \cite{Ketterle}.  
In this section, we use
the TBA approach to investigate the groundstate properties and
magnetism of the 1D mixture.  
Application of an external magnetic field
to the polarized fermions causes Zeeman splitting of the 
spin-up and spin-down fermions into different energy levels.  
The groundstate can only accommodate fermions that are in
the lower energy level. Therefore it is expected that when the
direction of the magnetic field is along the spin-up ($H>0$) direction,
spin-down fermions can no longer populate the groundstate. We
shall verify this picture from the TBA formalism.  
Under a magnetic field, the system possesses three phases: I) a pure boson phase; II)
a mixed boson-fermion phase; III) a fully polarized fermion phase.

The derivation of the TBA equations for the model (\ref{Ham-1}) is  standard \cite{Lai-2}. Here we present  a
proper set of dressed energy equations  which are convenient in the analysis  of
the groundstate properties.
These equations are
\begin{eqnarray}
\epsilon(k)&=&k^{2}-\mu-\frac12{H}+\frac{1}{2\pi}\int_{-B}^{B}\frac{c\phi^-(\lambda)}{\frac{c^2}{4}+(k-\lambda)^2}
\label{Dress0}\\
\phi(\lambda)&=&H+\frac{1}{2\pi}\int_{-Q}^{Q}\frac{c\epsilon^-(k)}{\frac{c^2}{4}+(\lambda-k)^2}\nonumber\\
& &+\, \frac{1}{2\pi}\int_{-d}^{d}\frac{c\psi^-(\Lambda)}{\frac{c^2}{4}+(\lambda-\Lambda)^2}-
\frac{1}{2\pi}\int_{-B}^{B}\frac{2c\phi^-(\lambda')}{c^2+(\lambda-\lambda')^2} \label{Dress1}
\\
\psi(\Lambda)&=&-\mu_{b}-\frac{H}{2}+\frac{1}{2\pi}\int_{-B}^{B}\frac{c\phi^-(\lambda)}{\frac{c^2}{4}+(\Lambda-\lambda)^2}. 
\label{Dress2}
\end{eqnarray}
The chemical potentials $\mu$ and $\mu_b$ are for the total
number of particles and for bosons.  
The magnetic field is again denoted by $H$. 
The dressed energies are defined \cite{Lai-2}
by $\exp(\epsilon (k)/T)=\rho^h(k)/\rho(k)$,
$\exp(\phi(\lambda)/T)=\sigma^h(\lambda)/\sigma(\lambda) $ and
$\exp(\psi (\Lambda)/T)=\tau^h(\Lambda)/\tau(\Lambda)$. Here $\rho(k)$
($\rho^h(k)$), $\sigma(\lambda)$ ($\sigma^h(\lambda)$) and
$\tau(\Lambda)$ ($\tau^h (\Lambda)$) are particle densities (hole
densities) in $k$ space, $\lambda $ space and $\Lambda$ space, determined
by the Bethe ansatz equations (\ref{BA-F}) in the thermodynamic limit
\cite{Lai-2}. The superscript $-$ denotes the negative part of the
dressed energies, which correspond to the occupied states. The
positive part of the dressed energies correspond to the unoccupied
states. The pressure of the system can be obtained from
\begin{equation}
P=\frac{1}{2\pi}\int_{-Q}^{Q}\epsilon^-(k)dk. \label{P}
\end{equation}

The dressed energy equations (\ref{Dress0})-(\ref{Dress2}) provide an
elegant way to analyze the groundstate properties in terms of
external fields. For $H=0$, we see that $\epsilon (k)$ has finite
Fermi points. However, $\phi(\lambda)$ and $\psi (\Lambda )$ have
Fermi surfaces at infinity. After taking Fourier transforms in equations (\ref{Dress1}) and (\ref{Dress2}) 
we obtain the single dressed energy equation
\begin{eqnarray}
\epsilon(k) = k^{2}-\mu-\mu_{b}+\frac{1}{2\pi}\int_{-Q}^{Q}\frac{2c\epsilon^-(k)}{c^2+(\lambda-k)^2}\label{g-e}
\end{eqnarray}
which is equivalent to the dressed energy equation for the spinless
Bose gas \cite{Y-Y,GBT} with the chemical potential $\mu_B=\mu+\mu_b$. 
Therefore, the groundstate for the mixture with
$H=0$ is exactly the same as that for  Lieb-Lininger bosons.  
For strong coupling, the groundstate energy 
\begin{equation}
\frac{E}{L}\approx \frac{1}{3}n^3\pi^2 \left(1-\frac{4}{\gamma}+\frac{12}{\gamma^2}\right)
\end{equation}
can be derived from (\ref{g-e}).
We have thus shown that the groundstate at $H=0$ is populated only by bosons.

As the external field is turned on, the Fermi surface for $\phi(\lambda)$ rises gradually. 
In this case, we have 
\begin{eqnarray}
\epsilon(k)&=&k^{2}-\mu-\frac12{H}+\frac{1}{2\pi}\int_{-B}^{B}\frac{c\phi^-(\lambda)}{\frac{c^2}{4}+(k-\lambda)^2}
\label{dress-1}
\\
\phi(\lambda)&=&-\mu_{b}+\frac12{H}+\frac{1}{2\pi}\int_{-Q}^{Q}\frac{c\epsilon^-(k)}{\frac{c^2}{4}+(\lambda-k)^2}. \label{dress-2}
\end{eqnarray}
Analysis of these dressed energy equations reveals that
pure Zeeman splitting between the two-component fermions does not 
energetically favour the fermions with spin-down. 
Therefore, the pure Zeeman field $H$ can only drive the system from the pure boson phase
into a mixture of fermions with spin-up and bosons or a phase of
fully polarized fermions for a large enough external field. 
The mixture of partially polarized fermions and bosons requires two Zeeman splitting
parameters which can maintain both fermions with spin-up and spin-down 
in certain regimes (see, e.g., the various regimes for 1D three-component interacting fermions \cite{GBLZ}).  
As the magnetic field is increased,
more fermions with spin-up populate the groundstate.
Beyond a critical  field value, the system will be entirely
occupied by fermions of one species. 
A phase transition is expected to occur when the field exceeds the critical value $H_{c}^M$.  
When $H\geq
H_{c}^M$, $\phi(\lambda)$ will become non-negative.  From the dressed
energy equations (\ref{dress-1}) and (\ref{dress-2}), we have
\begin{equation}
\phi(\lambda)\approx
-\mu_{b}+\frac12{H}-\frac{Pc}{c^{2}/4+\lambda^{2}}.
\end{equation}
Now when $H=H_{c}^M$, the Fermi sea for the spin pseudomomenta 
  vanishes, i.e., $\phi(0)=0$ and the density of bosons $m_b=M_b/L=0$.
Hence,
  the critical value is  $H_{c}^M={8P}/{c}$, where $P$ is the pressure
  per unit length of the system, which will be determined below.

For the mixture of fermions and bosons, the critical field is twice the
  value of that for spin-$\frac12$ fermions with a repulsive interaction
  discussed in the last section.  
The reason follows from the difference in their statistics.  
For the repulsive fermions, the groundstate
  is antiferromagnetic, with an  equal number of spin-up and spin-down fermions.
As the external magnetic field is increased it reaches the critical value $H_c^F$, at which 
half the total number of fermions with spin-down are flipped. 
Whereas in the mixture of polarized fermions and bosons the groundstate is populated only by bosons. 
When the field exceeds the critical value $H_c^M$, all of the bosons are driven out of the 
groundstate while all of the fermions are polarized. Therefore,
 full polarization in the mixture costs twice the energy of fully
  polarizing the spin-$\frac12$ fermions.

The dressed energy equations (\ref{dress-1}) and (\ref{dress-2}) can be dealt with 
analytically for an external field $0\le H\le H_{c}$ and in the strong coupling limit $\gamma \gg 1$.
Substitution of  equation (\ref{dress-2})  into equation (\ref{dress-1}) and some lengthy iteration calculations 
with the relations $\epsilon (Q)=0$ and $\phi(B)=0$ gives the leading behavior of the pressure per unit length as
\begin{eqnarray}
P &\approx& \frac{2\Omega^{3/2}}{3\pi}+\frac{H\Omega^{1/2}}{2\pi}\left[1-\frac{2}{\pi}\tan^{-1}
\left(\frac{2B}{c}\right)\right]\nonumber\\
&&+\frac{8\Omega^{2}}{3\pi^{3}c}\tan^{-1}\left(\frac{2B}{c}\right)
+\frac{16B\Omega^{2}}{3\pi^{3}(c^{2}+4B^{2})} \label{P-M}
\end{eqnarray}
where we have adopted the  notation 
\begin{equation}
\Omega=\mu+\frac{2\mu_{b}}{\pi}\tan^{-1}\left(\frac{2B}{c}\right).
\end{equation}
In the above equations the integration boundary for $\phi(\lambda)$ is
\begin{equation}
B\approx \frac{c}{2}\sqrt{\frac{8P/c-H+2\mu_{b}}{H-2\mu_{b}}}\label{B}
\end{equation}
and the Fermi point $Q$ can be expessed as 
\begin{eqnarray}
Q&\approx&\Omega^{1/2}\left\{1+
\frac{H}{4\Omega}\left[1-\frac{2}{\pi}\tan^{-1}\left(\frac{2B}{c}\right)\right]\right.\nonumber\\
&&\left.
+\frac{2P}{\Omega\pi c}\tan^{-1}\left(\frac{2B}{c}\right)+\frac{4PB}{\pi\Omega(c^{2}+4B^{2})}\right\}.
\end{eqnarray}

The density of bosons $m_b$ can be estimated from the pressure (\ref{P-M}) and the relation 
$m_b=\partial P/\partial \mu_b$ as
\begin{equation}
\frac{m_b}{n}\approx\frac{2}{\pi}\tan^{-1}\left(\frac{2B}{c}\right). \label{m}
\end{equation}
Using the relations (\ref{B}) and (\ref{m}) we have 
$\mu_b \approx  \frac12{H}-{4P c^{-1}\cos^2(\frac{m_b\pi}{2n})}$.  
The chemical
  potentials for bosons and fermions are  chosen as $\mu_B=\mu+\mu_b$ and $\mu_F =\mu$. 
Furthermore, on 
  neglecting terms of order $O(1/\gamma^2)$ and using the particle
  density relation $n=\partial P/\partial\mu$, we obtain the pressure
  per unit length and the  free energy in the form
\begin{eqnarray}
P &\approx &
\frac{2}{3}n^3\pi^2\left[1-\frac{6}{\gamma}\left(\frac{m_b}{n}+\frac{\sin(\frac{m_b\pi}{n})}{\pi}\right)\right] \\
F &\approx&
-\frac{(n-m_b)}{2}H+\frac{1}{3}\pi^{2}n^{3}\left[1-\frac{4}{\gamma}\left(\frac{m_b}{n}+\frac{\sin(\frac{m_b\pi}{n})}{\pi}\right)\right]
\end{eqnarray}
from which we can examine the magnetic properties of the model. 

\subsection{Magnetic properties}

We first see that the energy per unit length $E=F+m^zH$ is consistent with
the results obtained in Ref. \cite{BA1}. It follows that
\begin{equation}
H \approx \frac{8}{3\gamma}\pi^{2}n^{2}\left[1+\cos\left(\frac{m_b\pi}{n}\right)\right].\label{H-m}
\end{equation}
The susceptibility $\chi$ follows as
\begin{eqnarray}
 \chi (H)=\frac{\partial m^{z}}{\partial H} =\frac{3\gamma}{16\pi^{3}n\sin(\frac{m_b\pi}{n})}. \label{chi}
\end{eqnarray}
{}From the relation (\ref{H-m}), we find $\sin(\frac{m_b\pi}{n})=2\left(\frac{3\gamma
H}{16\pi^{2}n^{2}}\right)^{1/2}\left(1-\frac{3\gamma H}{16\pi^{2}n^{2}}\right)^{1/2}$.
As a result the susceptibility is given by 
\begin{eqnarray}
\chi &\approx& \frac{n}{2\pi}\frac{1}{H^{1/2}(H_{c}^M-H)^{1/2}}
\end{eqnarray}
where the critical field $H_{c}^M\approx \frac{16}{\gamma}E_F$.  
It is clear to see that the susceptibility is divergent in the vicinities of
$H=0$ and $H=H_c^M$. 
The magnetization in the vicinity of these points belongs to the universality class of
square-root field-dependent magnetization. 
This indicates a van Hove type singularity of the empty band. 
It is evident that pure Zeeman splitting for the mixture of polarized fermions and bosons 
may trigger three phases: a pure boson phase when the external field is absent; a fully polarized
phase when the external field exceeds the critical value $H_c^M$; and
a coexisting phase of fully polarized fermions and bosons for $0<H<H_{c}^M$, 
as shown in figure~\ref{fig:phase-m}.

This phase diagram is reminiscent of that of 1D weakly attractive fermions with population
imbalance \cite{Liu,HFGB}, where the fully paired phase occurs only
for the external field $H =0$; the fermions are fully polarized for $H \ge H_{c} = n^2(\pi^2+2|\gamma|)$; 
the paired and unpaired fermions coexist for $0<H<H_{c}$. 
However, there are subtle differences with the phase diagram for 
1D strongly attractive fermions with population imbalance \cite{GBLB}, 
where the bound pairs in the homogeneous system form a
singlet groundstate when the external field $H < H_{c1}$.
A completely ferromagnetic phase without pairing occurs when the
external field $H > H_{c2}$ and the paired and unpaired atoms coexist 
for an intermediate field $H_{c1} < H < H_{c2}$. 
The essential differences between the magnetic properties of the 
interacting Bose-Fermi mixture and interacting fermions with polarization are due to 
their different statistical signatures.

\begin{figure}
\includegraphics[width=0.82\textwidth]{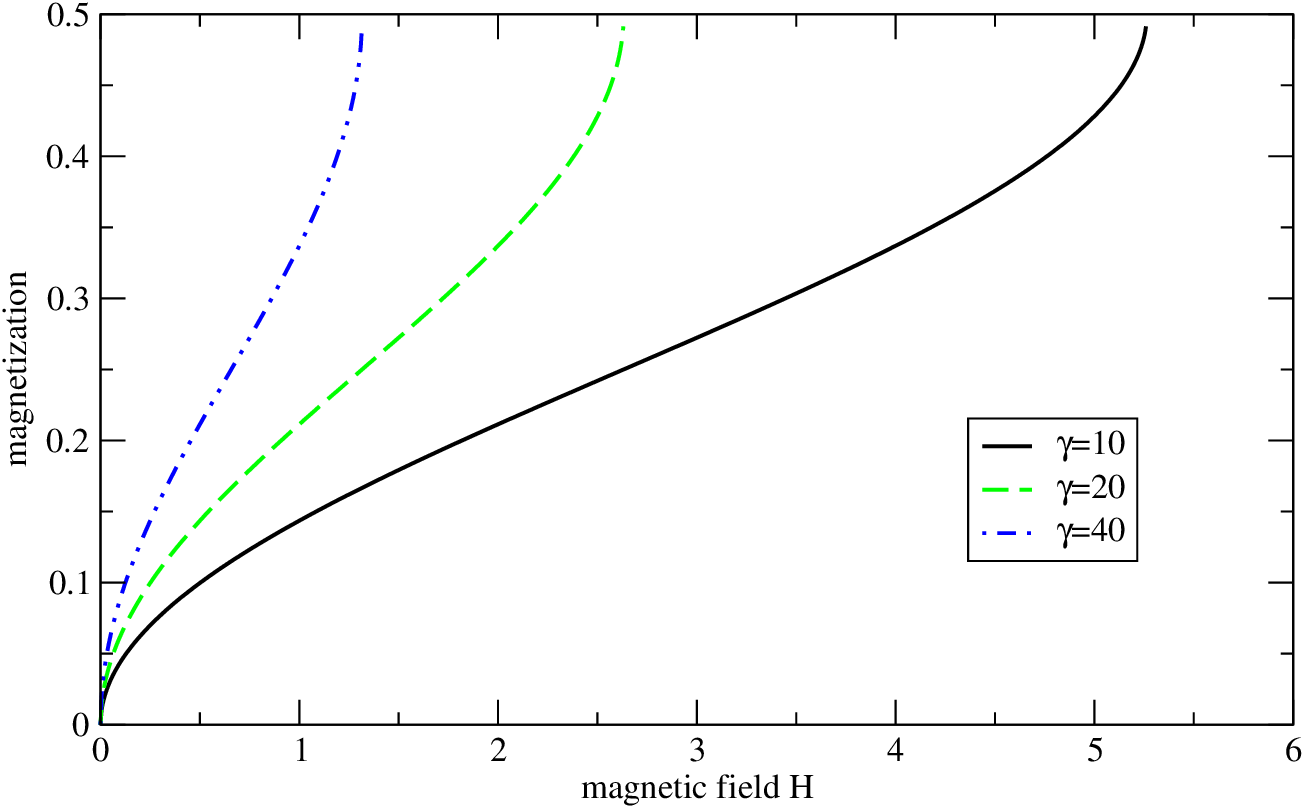}
\caption{Magnetization $m^{z}(H)$ vs
  external field $H$ for the Bose-Fermi mixture with $n=1$ and
  different values of $\gamma$. The
  magnetization curves are plotted from equation (\ref{H-m}).}  
\label{fig:mz}
\end{figure}

\begin{figure}
\includegraphics[width=0.82\textwidth]{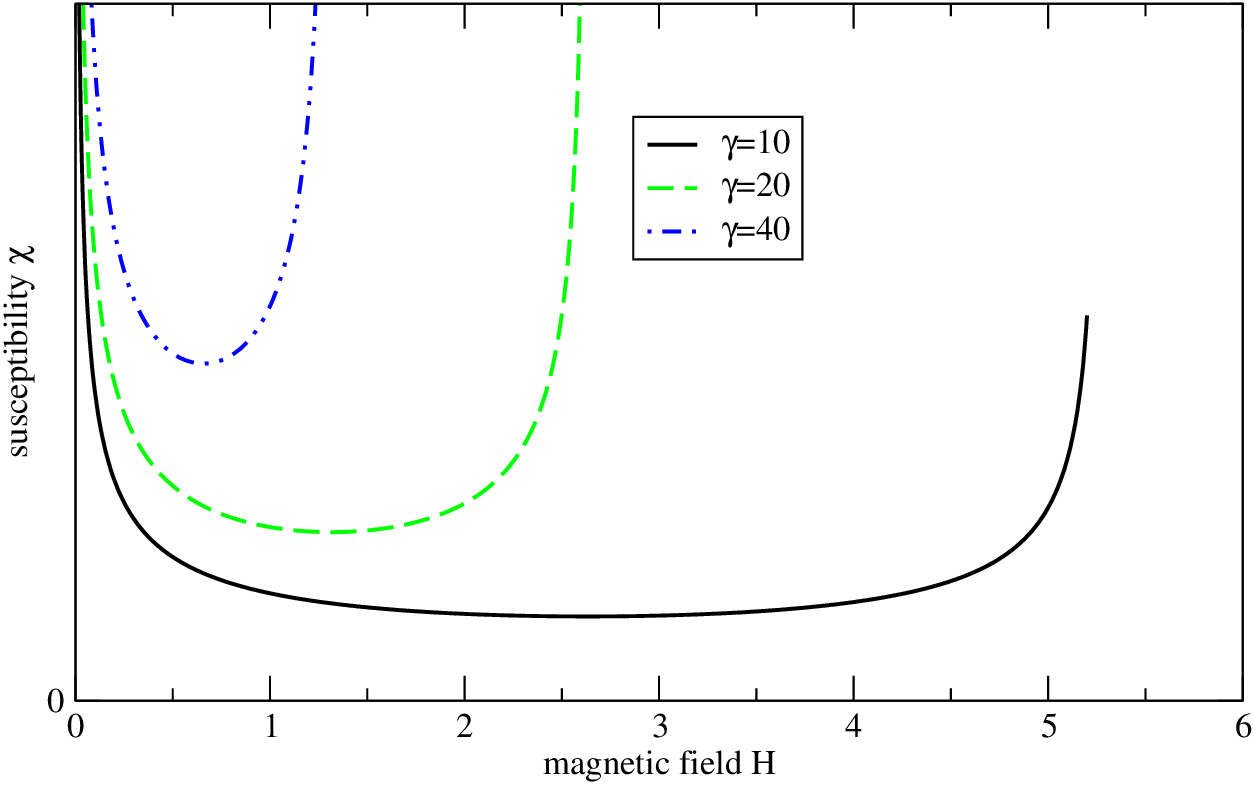}
\caption{Susceptibility $\chi(H)$ vs external field $H$ for the
  Bose-Fermi mixture with $n=1$ and different values of
  $\gamma$.  The susceptibility is evaluated
  from equation (\ref{chi}).}  
\label{fig:chi}
\end{figure}

Plots of  the magnetization and susceptibility versus the external magnetic field
$H$ are shown in figures~\ref{fig:mz} and \ref{fig:chi}. 
The susceptibility is infinitely divergent in the vicinities of $H=0$ and $H=H_c^M$. 
We can also see that as the interaction strength $\gamma$
increases the critical field tends to zero.  
If the interaction strength is weaker, the particles have more freedom to move along the line. 
Consequently, there will be a strong spin fluctuation which makes it harder to fully polarize 
the entire system. 
On the other hand, for strong repulsion the spins are ``frozen'', 
thus making it easier to fully polarize the system.

For the weak coupling regime, we obtain the groundstate
energy of the mixture of fully-polarized fermions and bosons from the Bethe ansatz 
equations (\ref{BA-M}), with result  \cite{BA2}
\begin{equation}
\frac{E}{L}\approx \left[\frac{M_1^3}{L^3}+\frac{M_b^2c}{L^2}+\frac{2M_bM_1c}{L^2}\right].
\end{equation}
The magnetization $m^z\approx \frac{1}{4\pi}(\sqrt{2H}+\frac{2c}{\pi})$ and  the susceptibility
$\chi \approx \frac{\sqrt{2}}{8\pi \sqrt{H}}$ follow from this equation. 
These results show that in the weak coupling regime the square-root field-dependent behavior of
magnetization emerges for finite external field. 
This is different from the antiferromagnetic behavior of the 1D Fermi gas with population imbalance.

\section{Conclusion}
\label{Conclusion}

\begin{figure}
\vskip 2mm
\includegraphics[width=0.82\textwidth]{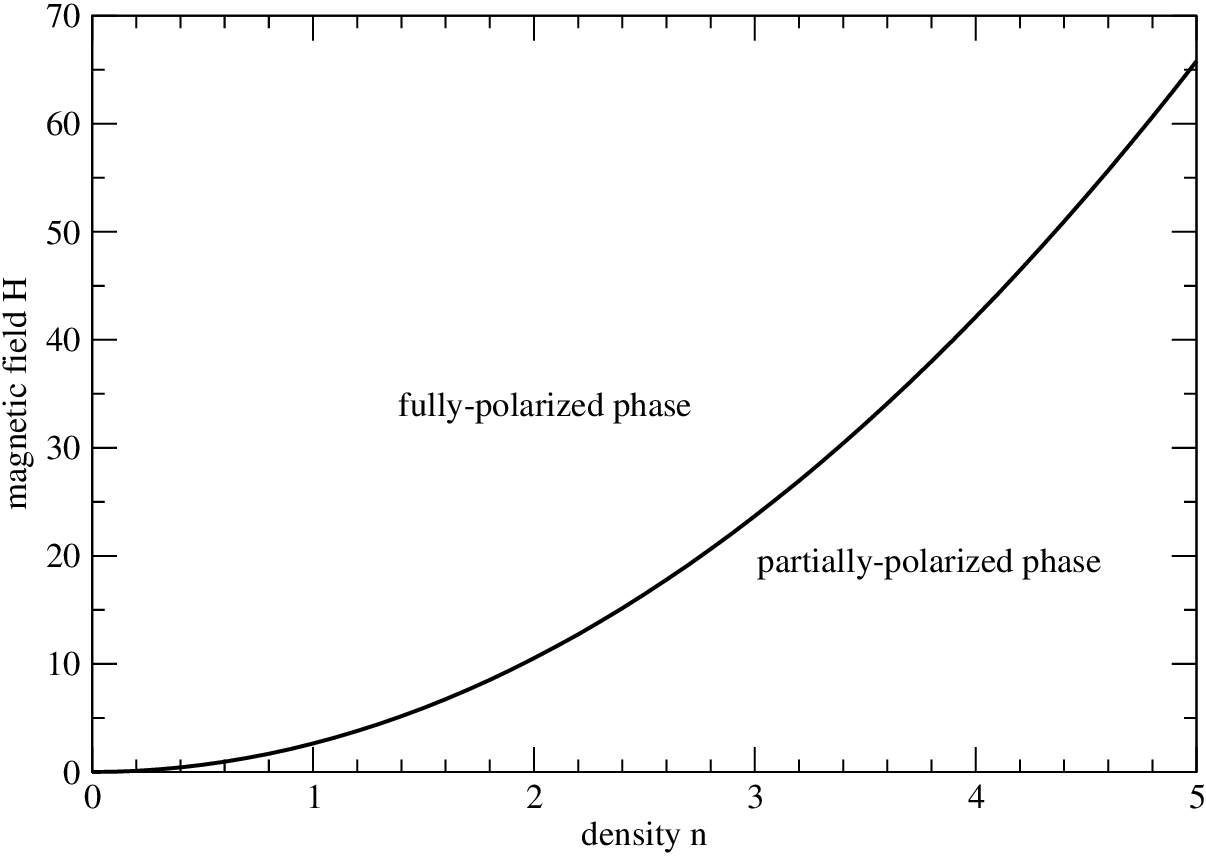}
\caption{Phase diagram for 1D polarized fermions with strong coupling $c=10$. }  
\label{fig:phase}
\end{figure}

We have studied external field-dependent  magnetic properties and statistical effects in 1D
polarized strongly repulsive fermions and in a 1D mixture of polarized
fermions and bosons with a repulsive interaction. 
We have found that the linear field-dependent behavior of the
magnetization in the polarized Fermi gas persists for weak and
finitely strong coupling regimes. The susceptibility is found to be 
finite. However, in the extreme limit $\gamma \to \infty$, the
susceptibility is infinitely divergent due to its paramagnetic
signature.  The spin-spin interaction is effectively
described by the isotropic  Heisenberg spin chain with antiferromagnetic coupling
constant $J\approx -4E_F/\gamma$. 
A quantum phase transition from the 
partially-polarized phase into the fully-polarized phase occurs when
the external field is greater than the critical value $H_c^F\approx 8E_F/\gamma$, 
recall figure~\ref{fig:phase}. For the weak coupling regime, the
critical field is  $H_c^F \approx n^2 (\pi^2 -2\gamma) $.  From these
configurations, we can predict the subtle segments for the 1D Fermi
gas in a harmonic trapping potential: partially-polarized fermions lie
in the center of the cloud whereas fully-polarized fermions sit in the
two outer wings (see also Ref.  \cite{MC}). 
The model of polarized fermions 
with repulsive interaction provides a tunable many-body system
exhibiting novel critical behavior. It is highly desirable to probe this 
many-body physics through experiments with 1D interacting fermions.

\begin{figure}
\includegraphics[width=0.82\textwidth]{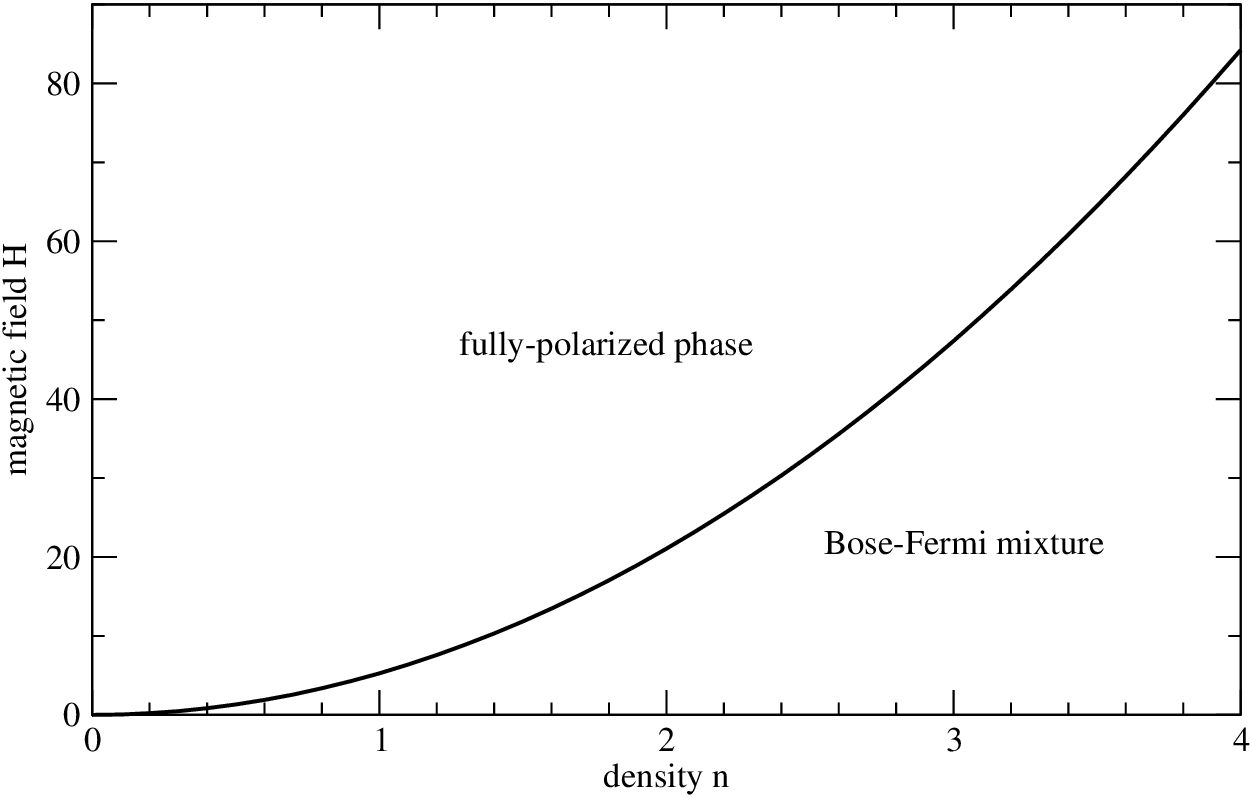}
\caption{Phase diagram for the 1D mixture of polarized fermions and
  bosons with strong coupling $c=10$.}
\label{fig:phase-m}
\end{figure}

For the mixture of polarized fermions and bosons we have shown that
the groundstate is only populated by bosons in the absence of a
magnetic field. When the external field is applied, spin-up fermions populate 
the groundstate. This leads to an infinitely divergent susceptibility. 
The fully-polarized
fermions and bosons coexist in the range $0<H<H_c^M\approx
16E_F/\gamma$. Another phase transition from this partially-polarized
phase into the fully-polarized Fermi phase takes place when the field
$H>H_c^M$, recall figure~\ref{fig:phase-m}. The susceptibility diverges. 
This phase diagram is somewhat reminiscent of weakly
interacting attractive fermions \cite{Liu}, where weakly interacting
BCS pairs can be viewed as strongly repulsive bosons in the mixture. 
This signature was recently observed in experiment
\cite{Ketterle}. There are subtle but essential differences between the mixture and the
attractive fermions due to the different statistical
signatures of the boson and the bosonic dimer. 
The mixture in a
harmonic trapping potential has distinct segments: a boson-fermion mixed 
phase in the center of the cloud and a fully-polarized fermion phase 
in the two outer wings (see also Ref. \cite{BA1}). These exotic
magnetic properties may also possibly be observed in experiment with
ultracold fermionic and bosonic atoms through  photoemission
spectroscopy techniques \cite{Jin}.

This work has been supported by the Australian Research Council.
The authors thank J.-P. Cao,  S. Chen and H.-Q. Zhou for helpful discussions.

\clearpage

\end{document}